\documentclass[11pt]{article}
\usepackage{amssymb,amsmath,amsfonts}
\usepackage{graphicx}
\usepackage{graphics}
\usepackage{eepic,epsfig}

\begin{document}

\title{Fermionic vacuum polarization by a cosmic string \\
in anti-de Sitter spacetime}
\author{E. R. Bezerra de Mello$^{1}$\thanks{%
E-mail: emello@fisica.ufpb.br}, E. R. Figueiredo Medeiros$^{1}$\thanks{%
E-mail: erfmedeiros@fisica.ufpb.br}, A. A. Saharian$^{1,2}$\thanks{%
E-mail: saharian@ysu.am} \\
%EndAName
\\
\textit{$^1$Departamento de F\'{\i}sica-CCEN, Universidade Federal da Para%
\'{\i}ba}\\
\textit{58.059-970, Caixa Postal 5.008, Jo\~{a}o Pessoa, PB, Brazil}\vspace{%
0.3cm}\\
\textit{$^2$Department of Physics, Yerevan State University,}\\
\textit{1 Alex Manoogian Street, 0025 Yerevan, Armenia}}
\maketitle

\begin{abstract}
In this paper we investigate the fermionic condensate (FC) and the vacuum
expectation value (VEV) of the energy-momentum tensor, associated with a
massive fermionic field, induced by the presence of a cosmic string in the
anti-de Sitter (AdS) spacetime. In order to develop this analysis we
construct the complete set of normalized eigenfunctions in the corresponding
spacetime. We consider a special case of boundary conditions on the AdS
boundary, when the MIT bag boundary condition is imposed on the field
operator at a finite distance from the boundary, which is then taken to
zero. The FC and the VEV of the energy-momentum tensor are decomposed into
the pure AdS and string-induced parts. Because the analysis of one-loop
quantum effects in the AdS spacetime has been developed in the literature,
here we are mainly interested to investigate the influence of the cosmic
string on the VEVs. The string-induced part in the VEV of the
energy-momentum tensor is diagonal and the axial and radial stresses are
equal to the energy density. For points near the string, the effects of the
curvature are subdominant and to leading order, the VEVs coincide with the
corresponding VEVs for the cosmic string in Minkowski bulk. At large proper
distances from the string, the decay of the VEVs show a power-law dependence
of the distance for both massless and massive fields. This is in contrast to
the case of Minkowski bulk where, for a massive field, the string-induced
parts decay exponentially.
\end{abstract}

\bigskip


\bigskip

\section{Introduction}

The nontrivial properties of the vacuum are among the most important
consequences in quantum field theory. These properties depend on both the
local and global geometrical characteristics of the background spacetime. In
particular, the boundary conditions imposed on a quantum field in spaces
with non-trivial topology modify the spectrum for zero-point fluctuations
which give rise to change in the vacuum expectation values (VEVs) for
physical observables. A well known example of this kind of phenomenon is the
topological Casimir effect. The explicit dependence of the physical
characteristics of the vacuum on the geometrical properties of the bulk, can
be found for highly-symmetric backgrounds only. In this paper we consider an
exactly solvable problem for the fermionic vacuum polarization by a cosmic
string in the background of anti-de Sitter (AdS) spacetime.

The importance of AdS spacetime as a background geometry in quantum field
theory is motivated by several reasons. The early interest was related to
principal questions of the quantization of fields on curved backgrounds. The
lack of global hyperbolicity and the presence of both regular and irregular
modes give rise to a number of new phenomena which have no analogs in
quantum field theory on the Minkowski bulk. These phenomena have been
discussed by several authors considering both scalar \cite{Avis}-\cite{Cald}
and spinor \cite{Camp1}-\cite{Eliz13} fields. The importance of this
research increased by the natural appearance of AdS spacetime as a ground
state in supergravity and Kaluza-Klein theories. The AdS geometry plays a
crucial role in two exciting developments in theoretical physics of the past
decade such as the AdS/CFT correspondence and the braneworld scenario with
large extra dimensions. The AdS/CFT correspondence (for a review see \cite%
{Ahar}), represents a realization of the holographic principle and relates
string theories or supergravity in AdS bulk with a conformal field theory
living on its boundary. The braneworld scenario offers a new perspective on
the hierarchy problem between the gravitational and electroweak mass scales
(for reviews on braneworld gravity and cosmology see \cite{Brax}). The main
idea to resolve the large hierarchy is that the small coupling of
4-dimensional gravity is generated by the large physical volume of extra
dimensions.

In the framework of grand unified theories, different types of topological
defects may have been created in the early universe as a consequence of the
vacuum phase transition \cite{Kibble,V-S}. Among them cosmic strings are of
special interest. Although recent observations data on the cosmic microwave
background have ruled out cosmic strings as the primary source for
primordial density perturbation, they are still candidate for the generation
of a number of interesting physical effects such as gamma ray burst \cite%
{Berezinski}, gravitational waves \cite{Damour} and high energy cosmic rays
\cite{Bhattacharjee}. Recently, cosmic strings have attracted renewed
interest partly because a variant of their formation mechanism is proposed
in the framework of brane inflation \cite{Sarangi}-\cite{Dvali}.

The geometry of a cosmic string in the background of AdS spacetime has been
considered in \cite{Ghe1,Cristine}. It has been shown that, similar to the
case of Minkowskian bulk, at distances from the string larger than its core
radius, the gravitational effects of the string can be described by a planar
angle deficit in the AdS line-element. The corresponding non-trivial
topology induces vacuum polarization effects for quantum fields. These
effects are well-investigated for the geometry of a cosmic string in
Minkowski spacetime for scalar \cite{scalar}-\cite{Mello1}, fermionic \cite%
{ferm}-\cite{Mello2} and vector quantum fields \cite{vector}-\cite{Mello3}.
Recently the corresponding results have been generalized for cosmic string
in curved spacetime. Specifically, one-loop quantum effects have been
considered for a massless scalar field in Schwarzschild spacetime \cite%
{Adrian}, for massive scalar \cite{String-dS} and fermionic \cite{String-dS1}
fields in de Sitter spacetime, and also for a massive scalar field in AdS
spacetime \cite{String-AdS}. In the present paper, we continue along similar
line of investigation, analyzing the fermionic vacuum polarization effects
associated with a massive Dirac spinor field in a four-dimensional AdS
spacetime in the presence of a cosmic string. Among the most important local
physical characteristics of the fermionic vacuum are the fermionic
condensate and the VEV of the energy-momentum tensor. The fermionic
condensate plays an important role in the models of dynamical breaking of
chiral symmetry, whereas the VEV of the energy-momentum tensor acts as the
source of gravity in the quasiclassical Einstein equations.

This paper is organized as follows: In section \ref{sec2} we present the
background geometry associated with the spacetime under consideration and
construct the complete set of positive- and negative-energy fermionic
wave-functions. In section \ref{sec3} we calculate the fermion condensate by
using the mode-summation method. In this evaluation the contribution induced
by the cosmic string is separately analyzed and its behavior in the
asymptotic regions of the parameters is investigated. The analysis of the
VEV of the energy-momentum tensor is considered in section \ref{sec4}.
There, the contribution induced by the cosmic string is also investigated.
Finally, the main results of the paper are summarized in section \ref{conc}.
In appendix we show that the mode functions used in the main text are
obtained when the MIT bag boundary condition is imposed at a finite distance
from the AdS boundary, which is then taken to zero. Throughout of the paper
we shall use the units $\hbar =c=G=1$.

\section{Fermionic wave-functions}

\label{sec2}

The main objective of this section is to obtain the complete set of
fermionic mode-functions in a four-dimensional AdS spacetime in presence of
a cosmic string. This set is needed in the calculation of vacuum
polarization effects by using the mode-summation approach.

In cylindrical coordinates, considering a static string along the $y$-axis,
the geometry associated with a cosmic string in a four-dimensional AdS
spacetime is given by the following line element:
\begin{equation}
ds^{2}=e^{-2y/a}\left( -dt^{2}+dr^{2}+r^{2}d\phi ^{2}\right) +dy^{2}\ ,
\label{ds1}
\end{equation}%
where $r\geq 0$ and $\phi \in \lbrack 0,\ 2\pi /q]$ define the coordinates
on the conical subspace, $(t,\ y)\in (-\infty ,\ \infty )$. The points $(r,\
\phi ,\ z)$ and $(r,\phi +2\pi /q,z)$ are to be identified, and the
parameter $a$ is related with the cosmological constant and Ricci scalar for
AdS spacetime by the formulas
\begin{equation}
\Lambda =-3a^{-2},\ R=-12a^{-2}\ .  \label{R}
\end{equation}%
The parameter $q$, bigger than unity, codifies the presence of the cosmic
string.

By using the Poincar\'{e} coordinate defined as $z=ae^{y/a}$, the line
element above is presented in the form conformally related to the line
element associated with a cosmic string in Minkowski spacetime:
\begin{equation}
ds^{2}=(a/z)^{2}\left( -dt^{2}+dr^{2}+r^{2}d\phi ^{2}+dz^{2}\right) \ .
\label{ds2}
\end{equation}%
For this new coordinate one has $z\in \lbrack 0,\ \infty )$. The
hypersurfaces $z=0$ and $z=\infty $ correspond to the AdS boundary and
horizon, respectively.

Although the line element for an infinite straight cosmic string in the
background of Minkowski spacetime, which corresponds to the expression
inside the parentheses in (\ref{ds2}), has been derived in \cite{Vile81} by
making use of weak-field approximation, the validity of this solution has
been extended beyond the linear perturbation theory by several authors \cite%
{Gott85}. In this case the parameter $q$ need not to be close to the unity.
The metric tensor corresponding to the line element (\ref{ds1}) is an exact
solution of the Einstein equation in the presence of a negative cosmological
constant and the string \cite{Ghe1,Cristine}, for arbitrary values of $q$.

The quantum motion of a massive spinor field on curved spacetime is governed
by the Dirac equation
\begin{equation}
i\gamma ^{\mu }\nabla _{\mu }\psi -m\psi =0\ ,\;\nabla _{\mu }=\partial
_{\mu }+\Gamma _{\mu }\ ,  \label{Direq}
\end{equation}%
where $\gamma ^{\mu }$ are the Dirac matrices in curved spacetime and $%
\Gamma _{\mu }$ is the spin connection. They are given in terms of the flat
spacetime Dirac matrices, $\gamma ^{(a)}$, by the relations,
\begin{equation}
\gamma ^{\mu }=e_{(a)}^{\mu }\gamma ^{(a)}\ ,\ \Gamma _{\mu }=-\frac{1}{4}%
\gamma ^{(a)}\gamma ^{(b)}e_{(a)}^{\nu }e_{(b)\nu ;\mu }\ ,  \label{Gammamu}
\end{equation}%
where the semicolon means the standard covariant derivative for vector
fields. In (\ref{Gammamu}), $e_{(a)}^{\mu }$ is the tetrad basis satisfying
the relation $e_{(a)}^{\mu }e_{(b)}^{\nu }\eta ^{ab}=g^{\mu \nu }$, with $%
\eta ^{ab}$ being the Minkowski spacetime metric tensor.

In order to simplify the obtainment of the fermionic wave-functions in the
geometry under consideration, we shall use for the flat space Dirac matrices
the representation which is obtained from the matrices given in \cite{Bjor64}
multiplied on the left by $i\gamma ^{(5)}$ matrix. In this representation
one has
\begin{equation}
\gamma ^{(0)}=-i\left(
\begin{array}{cc}
0 & 1 \\
-1 & 0%
\end{array}%
\right) \ ,\ \gamma ^{(a)}=-i\left(
\begin{array}{cc}
\sigma _{a} & 0 \\
0 & -\sigma _{a}%
\end{array}%
\right) \ ,\ a=1,2,3\ ,  \label{gam0l}
\end{equation}%
with $\sigma _{1},\sigma _{2},\sigma _{3}$ being the Pauli matrices. It can
be verified that the above matrices obey the Clifford algebra, $\{\gamma
^{(a)},\ \gamma ^{(b)}\}=-2\eta ^{ab}$.

The basis of tetrads corresponding to the line element (\ref{ds2}) can be
taken in the form
\begin{equation}
e_{(a)}^{\mu }=\frac{z}{a}\left(
\begin{array}{cccc}
1 & 0 & 0 & 0 \\
0 & \cos (q\phi ) & -\sin (q\phi )/r & 0 \\
0 & \sin (q\phi ) & \cos (q\phi )/r & 0 \\
0 & 0 & 0 & 1%
\end{array}%
\right) \ .  \label{Tetrad}
\end{equation}%
With this choice, the curved space gamma matrices read:
\begin{equation}
\gamma ^{0}=\frac{z}{a}\gamma ^{(0)}\ ,\ \gamma ^{i}=-i\frac{z}{a}\left(
\begin{array}{cc}
\sigma ^{i} & 0 \\
0 & -\sigma ^{i}%
\end{array}%
\right) \ ,  \label{gam02}
\end{equation}%
where the index $i$ corresponds to the coordinates $r,\ \phi ,\ z$. The $%
2\times 2$ matrices in (\ref{gam02}) are given by
\begin{equation}
\sigma ^{r}=\left(
\begin{array}{cc}
0 & e^{-iq\phi } \\
e^{iq\phi } & 0%
\end{array}%
\right) ,\ \sigma ^{\phi }=-\frac{i}{r}\left(
\begin{array}{cc}
0 & e^{-iq\phi } \\
-e^{iq\phi } & 0%
\end{array}%
\right) ,\ \ \sigma ^{z}=\left(
\begin{array}{cc}
1 & 0 \\
0 & -1%
\end{array}%
\right) .  \label{betal}
\end{equation}%
For the spin connection components we obtain:
\begin{equation}
\Gamma _{\mu }=-\frac{1}{2a}\gamma ^{(3)}\gamma _{\mu }+\frac{(1-q)}{2}%
\gamma ^{(1)}\gamma ^{(2)}\delta _{\mu }^{\phi }\ ,\ \Gamma _{z}=0\ .
\label{Gaml}
\end{equation}%
This leads to the following expression for the combination appearing in the
Dirac equation (\ref{Direq}):
\begin{equation}
\gamma ^{\mu }\Gamma _{\mu }=-\frac{3}{2z}\gamma ^{z}+\frac{1-q}{2r}\gamma
^{r}\ .  \label{gamGam}
\end{equation}

Taking the time dependence for the positive-energy four-component spinor in
the form $e^{-iEt}$, and decomposing it into upper and lower two-component
spinors denoted by $\varphi $ and $\chi $, respectively, the Dirac equation
can be written as shown below:
\begin{eqnarray}
\left( \sigma ^{l}\partial _{l}-\frac{3}{2z}\sigma ^{z}+\frac{1-q}{2r}\sigma
^{r}-\frac{ma}{z}\right) \varphi -iE\chi &=&0\ ,  \notag \\
\left( \sigma ^{l}\partial _{l}-\frac{3}{2z}\sigma ^{z}+\frac{1-q}{2r}\sigma
^{r}+\frac{ma}{z}\right) \chi -iE\varphi &=&0\ .  \label{Dirac1}
\end{eqnarray}%
Substituting the spinor $\chi $ from the first equation into the second one,
we obtain a second order differential equation for $\varphi $:
\begin{eqnarray}
&&\left[ \partial _{r}^{2}+\frac{1}{r}\partial _{r}-i\frac{1-q}{r^{2}}\sigma
^{3}\partial _{\phi }+\frac{1}{r^{2}}\partial _{\phi }^{2}+\partial _{z}^{2}-%
\frac{3}{z}\partial _{z}\right.  \notag \\
&&\left. -\frac{(1-q)^{2}}{4r^{2}}+\frac{15-4m^{2}a^{2}}{4z^{2}}+\frac{ma}{%
z^{2}}\sigma ^{z}+E^{2}\right] \varphi =0\ .  \label{phieq}
\end{eqnarray}%
Because the above equation has a diagonal form, we can obtain two second
order differential equations decomposing the two-component spinor $\varphi $
into the upper and lower components, denoted by $\varphi _{1}$ and $\varphi
_{2}$, respectively. Taking these functions in the form $\varphi
_{l}(x)=e^{iqn_{l}\phi }R_{l}(r)Z_{l}(z)$, with $l=1,2$, we can see that the
normalizable solution for the radial function is expressed in terms of the
Bessel function of the first kind, $R_{l}(r)=J_{\beta _{l}}(\lambda r)$,
with the order,
\begin{equation}
\beta _{l}=|qn_{l}-(-1)^{l}(q-1)/2|,  \label{betl}
\end{equation}%
where $n_{l}=0,\pm 1,\pm 2,\ ...$. As to the function $Z_{l}(z)$, the
general solution is given in terms of a linear combination of the functions $%
z^{2}J_{\nu _{l}}(kz)$ and $z^{2}Y_{\nu _{l}}(kz)$, where $Y_{\nu }(x)$ is
the Neumann function and
\begin{equation}
\nu _{l}=ma+(-1)^{l}/2\ .  \label{nul}
\end{equation}%
For $ma\geq 1/2$ the part with the Neumann function is excluded by the
normalizability condition of the mode functions and the solution for the
function $Z_{l}(z)$ reads:
\begin{equation}
Z_{l}(z)=z^{2}J_{\nu _{l}}(kz)\ .  \label{Zl}
\end{equation}%
As to the energy, it is given by
\begin{equation}
E=\sqrt{\lambda ^{2}+k^{2}}\ .  \label{Ek}
\end{equation}

For $ma<1/2$, the modes with the Neumann function are normalizable as well.
In this case, in order to specify uniquely the mode functions, an additional
boundary condition is required on the AdS boundary $z=0$. Here, for the case
$ma<1/2$, we consider a special case of boundary conditions at the AdS
boundary, when the MIT bag boundary condition is imposed at a finite
distance from the boundary, $z=\delta $, which is then taken to zero, $%
\delta \rightarrow 0$ (FC and the VEV of the energy-momentum tensor for the
geometry of two flat boundaries in AdS bulk with the bag boundary conditions
have been recently discussed in \cite{Eliz13}). In Appendix we show that
this procedure leads to the choice of the mode functions in the form (\ref%
{Zl}) and ensures the zero current of fermions through the AdS boundary.
Note that the boundary condition we have used also excludes the normalizable
modes with $Z_{l}(z)=z^{2}K_{\nu _{l}}(z\sqrt{\lambda ^{2}-E^{2}})$, $%
\lambda >E$, where $K_{\nu }(x)$ is the Macdonald function.

Finally, we can write the upper two-component spinor in the form
\begin{equation}
\varphi _{l}=C_{l}z^{2}J_{\beta _{l}}(\lambda r)J_{\nu
_{l}}(kz)e^{iqn_{l}\phi }\ .  \label{upper}
\end{equation}%
Having the upper component spinor, the lower one is obtained by using the
first equation of (\ref{Dirac1}). After some intermediate steps we find:
\begin{eqnarray}
\chi _{1} &=&B_{1}e^{iqn_{1}\phi }J_{\beta _{1}}(\lambda r)z^{2}J_{\nu
_{2}}(kz)\ ,  \notag \\
\chi _{2} &=&B_{2}e^{iqn_{2}\phi }J_{\beta _{2}}(\lambda r)z^{2}J_{\nu
_{1}}(kz)\ ,  \label{lower1}
\end{eqnarray}%
with the relations
\begin{equation}
n_{2}=n_{1}+1\ ,\ \beta _{2}=\beta _{1}+\epsilon _{n_{1}}\ ,  \label{rel12}
\end{equation}%
where $\epsilon _{n}=1$ for $n\geq 0$ and $\epsilon _{n}=-1$ for $n<0$. The
coefficients $B_{1,2}$ in (\ref{lower1}) are given by the expressions
\begin{equation}
B_{1}=\frac{i}{E}\left( kC_{1}-\epsilon _{n_{1}}\lambda C_{2}\right) \ ,\
B_{2}=\frac{i}{E}\left( \epsilon _{n_{1}}\lambda C_{1}+kC_{2}\right) \ .
\label{B12}
\end{equation}

We can see that the fermionic wave-functions defined with the upper and
lower components given by relations (\ref{upper}) and (\ref{lower1}),
respectively, are eigenfunctions of the projection of the total momentum
along the direction of the cosmic string:
\begin{equation}
\widehat{J}_{z}\psi =\left( -i\partial _{\phi }+\frac{q}{2}\Sigma
^{z}\right) \psi =qj\psi \ ,\ \Sigma ^{z}=\left(
\begin{array}{cc}
\sigma ^{z} & 0 \\
0 & \sigma ^{z}%
\end{array}%
\right) \ ,  \label{J3}
\end{equation}%
where
\begin{equation}
j=n_{1}+1/2\ ,\ j=\pm 1/2,\pm 3/2\ ,\ldots \ .  \label{j}
\end{equation}

The fermionic wave-functions we have derived contain four coefficients and
there are two equations relating them. The normalization condition on the
functions provides an extra equation. Consequently, one of the coefficients
remains arbitrary. In order to determine this coefficient some additional
condition should be imposed on the coefficients. The necessity for this
condition is related to the fact that the quantum numbers $(\lambda ,k,j)$
do not specify the fermionic wave-function uniquely and some additional
quantum number is required.

In order to specify the second constant we impose the condition
\begin{equation}
C_{1}/B_{1}=-C_{2}/B_{2}\ .  \label{AdCond}
\end{equation}%
By taking into account (\ref{B12}) we get the relations:
\begin{equation}
C_{2}=-\epsilon _{n_{1}}b_{s}^{(+)}C_{1}\ ,\ B_{1}=isC_{1}\ ,\
B_{2}=i\epsilon _{n_{1}}sb_{s}^{(+)}C_{1}\ ,  \label{CoefRel}
\end{equation}%
where and in what follows
\begin{equation}
b_{s}^{(\pm )}=\frac{sE\mp k}{\lambda }\ ,\ \ s=\pm 1\ .  \label{bsplm}
\end{equation}%
Note that one has $b_{s}^{(+)}b_{s}^{(-)}=1$. With the condition (\ref%
{AdCond}), the fermionic mode functions are uniquely specified by the set $%
\sigma =(\lambda ,k,j,s)$. Instead of (\ref{AdCond}) we could impose another
condition. The only restriction is that the resulting wave-functions should
form a complete set. For example, the condition similar to (\ref{AdCond})
with the opposite sign of the right-hand side gives another set of
wave-functions. Different conditions give the same two-point functions and,
as a result, the same VEVs for physical observables.

On the basis of the discussion above, for the positive-energy fermionic
wave-function we can write the following expression:
\begin{equation}
\psi _{\sigma }^{(+)}(x)=C_{\sigma }^{(+)}e^{-iEt}z^{2}\left(
\begin{array}{c}
J_{\beta _{1}}(\lambda r)J_{\nu _{1}}(kz) \\
-\epsilon _{j}b_{s}^{(+)}J_{\beta _{2}}(\lambda r)J_{\nu _{2}}(kz)e^{iq\phi }
\\
isJ_{\beta _{1}}(\lambda r)J_{\nu _{2}}(kz) \\
i\epsilon _{j}sb_{s}^{(+)}J_{\beta _{2}}(\lambda r)J_{\nu _{1}}(kz)e^{iq\phi
}%
\end{array}%
\right) e^{iq(j-1/2)\phi }\ ,  \label{psi+}
\end{equation}%
where the order of the Bessel functions are defined in terms of $j$ as:
\begin{equation}
\beta _{l}=|qj+(-1)^{l}/2|=q|j|+(-1)^{l}\epsilon _{j}/2\ .  \label{bet12a}
\end{equation}%
Notice that $\epsilon _{j}=\epsilon _{n_{1}}$.

The coefficient $C_{\sigma }^{(+)}$ in (\ref{psi+}) is determined from the
normalization condition
\begin{equation}
\int d^{3}x\sqrt{\gamma }\ (\psi _{\sigma }^{(+)})^{\dagger }\psi _{\sigma
^{\prime }}^{(+)}=\delta _{\sigma \sigma ^{\prime }}\ ,  \label{normcond}
\end{equation}%
where $\gamma $ is the determinant of the spatial metric. The delta symbol
on the right-hand side is understood as the Dirac delta function for
continuous quantum numbers $(\lambda ,\ k)$ and the Kronecker delta for
discrete ones $(j,s)$. Substituting the eigenspinors (\ref{psi+}) into (\ref%
{normcond}) and using the value of the standard integral involving the
products of the Bessel functions \cite{Grad}, we find
\begin{equation}
|C_{\sigma }^{(+)}|^{2}=\frac{sq\lambda ^{2}k}{8\pi a^{3}Eb_{s}^{(+)}}\ .
\label{coef+}
\end{equation}

In the discussion above, as solutions of the equations for the radial parts
of the mode functions we have taken the Bessel functions $J_{\beta
_{l}}(\lambda r)$, $l=1,2$. This choice corresponds to the modes regular on
the string. In addition, we could consider the irregular modes with the
Neumann functions $Y_{\beta _{l}}(\lambda r)$. For these modes to be
normalizable the integral $\int_{0}^{\infty }dr\,rY_{\beta _{l}}(\lambda
r)Y_{\beta _{l}}(\lambda ^{\prime }r)$ should converge at the lower limit of
the integration for both $l=1$ and $l=2$. From the condition of the
convergence for $l=2$ one gets the constraint $q|j|<1/2$, which cannot be
satisfied for $q\geqslant 1$. Hence, in the problem under consideration
there are no normalizable modes irregular on the string. Note that,
irregular normalizable modes are possible in the presence of a magnetic flux
running along the axis of the string.

The negative-energy fermionic wave-function, can be obtained from the
positive-energy function by the charge conjugate matrix. Following the
general procedure given in \cite{Bjor64}, the charge conjugation matrix, $%
\mathcal{C}$, must obey the relation ${\mathcal{C}}(\gamma ^{(a)})^{\ast }{%
\mathcal{C}}^{-1}=-\gamma ^{(a)}$, with $(\gamma ^{(a)})^{\ast }$ denoting
the complex conjugate. In the representation used by us for the Dirac
matrices, all but $\gamma ^{(2)}$ matrix are complex ones; so $\mathcal{C}$
must anti-commute with $\gamma ^{(2)}$ and commute with the others. Up to an
overall phase, a particular choice for this matrix is,
\begin{equation*}
{\mathcal{C}}=\gamma ^{(5)}\gamma ^{(2)}=i\left(
\begin{array}{cc}
0 & \sigma _{2} \\
-\sigma _{2} & 0%
\end{array}%
\right) ={\mathcal{C}}^{-1}\ .
\end{equation*}%
The negative-energy wave-function can be given by $\psi _{\sigma }^{(-)}(x)={%
\mathcal{C}}(\psi _{\sigma }^{(+)}(x))^{\ast }$. The final result is
\begin{equation}
\psi _{\sigma }^{(-)}(x)=C_{\sigma }^{(-)}e^{iEt}z^{2}\left(
\begin{array}{c}
J_{\beta _{2}}(\lambda r)J_{\nu _{1}}(kz) \\
-\epsilon _{j}b_{s}^{(-)}J_{\beta _{1}}(\lambda r)J_{\nu _{2}}(kz)e^{iq\phi }
\\
isJ_{\beta _{2}}(\lambda r)J_{\nu _{2}}(kz) \\
i\epsilon _{j}sb_{s}^{(-)}J_{\beta _{1}}(\lambda r)J_{\nu _{1}}(kz)e^{iq\phi
}%
\end{array}%
\right) e^{-iq(j+1/2)\phi }\ ,  \label{psi-}
\end{equation}%
with the same notations as in (\ref{psi+}), and $C_{\sigma
}^{(-)}=-i\epsilon _{j}sb_{s}^{(+)}(C_{\sigma }^{(+)})^{\ast }$. The modulus
of the normalization constant $C_{\sigma }^{(-)}$ can also be evaluated by
using the normalization condition similar to the one for the positive-energy
wave-function, Eq. (\ref{normcond}). Its value reads,
\begin{equation}
|C_{\sigma }^{(-)}|^{2}=\frac{sq\lambda ^{2}k}{8\pi a^{3}Eb_{s}^{(-)}}\ .
\label{coef-}
\end{equation}%
The wave-functions obtained in this section can be used for the
investigation of various quantum effects around the cosmic string involving
electrons and positrons. In what follows we use these functions for the
evaluation of the fermionic condensate and the VEV of the energy-momentum
tensor.

\section{Fermionic condensate}

\label{sec3}

Having obtained the normalized positive- and negative-energy fermionic
wave-functions, we are able to calculate the fermionic condensate (FC), $%
\langle 0|\bar{\psi}\psi |0\rangle $, where $|0\rangle $ is the vacuum state
in the AdS spacetime in the presence of a cosmic string, and $\bar{\psi}%
_{\sigma }^{(-)}=\psi _{\sigma }^{(-)\dagger }\gamma ^{(0)}$ is the Dirac
adjoint. Expanding the field operator in terms of the complete set $\{\psi
_{\sigma }^{(-)},\ \psi _{\sigma }^{(+)}\}$, the following formula for the
FC is obtained:
\begin{equation}
\langle 0|\bar{\psi}\psi |0\rangle =\sum_{\sigma }\bar{\psi}_{\sigma
}^{(-)}\psi _{\sigma }^{(-)}\ ,  \label{FCmodesum}
\end{equation}%
where we use the compact notation defined below,
\begin{equation}
\sum_{\sigma }=\int_{0}^{\infty }\ d\lambda \ \int_{0}^{\infty }\ dk\
\sum_{s=\pm 1}\sum_{j=\pm 1/2,\cdots }\ .  \label{Sumsig}
\end{equation}

Substituting the eigenspinor (\ref{psi-}) into (\ref{FCmodesum}), we obtain
\begin{eqnarray}
\langle 0|\bar{\psi}\psi |0\rangle &=&\frac{qz^{4}}{4\pi a^{3}}\sum_{\sigma }%
\frac{\lambda ^{2}\ k}{E}J_{\nu _{1}}(kz)J_{\nu _{2}}(kz)  \notag \\
&&\times \left[ b_{s}^{(+)}J_{\beta _{2}}^{2}(\lambda r)-b_{s}^{(-)}J_{\beta
_{1}}^{2}(\lambda r)\right] \ .  \label{FC1}
\end{eqnarray}%
The fermionic condensate given by (\ref{FC1}) is divergent and some
regularization procedure is necessary. We shall assume that a cutoff
function is introduced in the above formula without explicitly writing it.
As we shall see, the explicit form of this function is not relevant for the
further discussion

The sum over the total angular quantum number $j$ for the corresponding
Bessel functions provide the same expression for $l=1$ and $l=2$:
\begin{equation}
\sum_{j=\pm 1/2,\cdots }J_{\beta _{l}}^{2}(\lambda r)=\sum_{j}\left[
J_{qj-1/2}^{2}(\lambda r)+J_{qj+1/2}^{2}(\lambda r)\right] \ .  \label{sum-j}
\end{equation}%
Here and in what follows, we use the notation
\begin{equation}
\sum_{j}=\sum_{j=1/2,\cdots }\ .  \label{SumjNot}
\end{equation}%
Now it is possible to develop the summation over the quantum number $s$. The
result is expressed as,
\begin{equation}
\langle 0|\bar{\psi}\psi |0\rangle =-\frac{q\ z^{4}}{\pi a^{3}}%
\int_{0}^{\infty }d\lambda \int_{0}^{\infty }dk\frac{\lambda \ k^{2}}{E}%
J_{\nu _{1}}(kz)J_{\nu _{2}}(kz)\sum_{j}\left[ J_{qj+1/2}^{2}(\lambda
r)+J_{qj-1/2}^{2}(\lambda r)\right] \ .  \label{FC2}
\end{equation}%
In order to provide a more workable expression for the FC, we use the
identity
\begin{equation}
\frac{1}{E}=\frac{2}{\sqrt{\pi }}\int_{0}^{\infty }ds\ e^{-(\lambda
^{2}+k^{2})s^{2}}\ .  \label{ident1}
\end{equation}%
Substituting this identity into (\ref{FC2}), it is possible to develop the
integrations over the variables $\lambda $ and $k$ with the help of formula
from \cite{Grad}. Although the orders of the Bessel functions associated
with the $z$-variable are different, the integral over $k$ is evaluated
because $\nu _{2}=\nu _{1}+1$. By using the relation, $J_{\nu }(x)J_{\nu
+1}(x)=(\nu /x)J_{\nu }^{2}(x)-\partial _{x}J_{\nu }^{2}(x)/2$, after some
intermediate steps the integral reads,
\begin{equation}
\int_{0}^{\infty }dk\,k^{2}e^{-k^{2}s^{2}}J_{\nu _{1}}(kz)J_{\nu _{2}}(kz)=%
\frac{e^{-z^{2}/(2s^{2})}}{4s^{4}}z\left[ I_{\nu
_{1}}(z^{2}/(2s^{2}))-I_{\nu _{2}}(z^{2}/(2s^{2}))\right] \ .  \label{Int1}
\end{equation}%
Now, introducing a new variable $y=r^{2}/(2s^{2})$, we obtain
\begin{equation}
\langle 0|\bar{\psi}\psi |0\rangle =-\frac{q\rho ^{-5}}{\sqrt{2\pi }\pi a^{3}%
}\int_{0}^{\infty }dy\ y^{3/2}e^{-(1+\rho ^{-2})y}\left[ I_{\nu _{1}}\left(
y/\rho ^{2}\right) -I_{\nu _{2}}\left( y/\rho ^{2}\right) \right] \mathcal{I}%
(q,y)\ ,  \label{FC3}
\end{equation}%
with the notations%
\begin{equation}
\rho =r/z\ ,  \label{rho}
\end{equation}%
and%
\begin{equation}
\mathcal{I}(q,y)=\sum_{j}\left[ I_{qj-1/2}(y)+I_{qj+1/2}(y)\right] .
\label{Iqy}
\end{equation}

First we consider the FC for a massless field. In this case, from (\ref{FC3}%
) one finds
\begin{equation}
\langle 0|\bar{\psi}\psi |0\rangle =\left( \frac{z}{a}\right) ^{3}\langle
\bar{\psi}\psi \rangle _{\mathrm{b}}^{\mathrm{(M)}},  \label{FCm0}
\end{equation}%
where%
\begin{equation}
\langle \bar{\psi}\psi \rangle _{\mathrm{b}}^{\mathrm{(M)}}=-\frac{qz}{\pi
^{2}r^{4}}\int_{0}^{\infty }dy\,ye^{-(2z^{2}/r^{2}+1)y}\mathcal{I}(q,y),
\label{FCb3}
\end{equation}%
is the FC for a massless fermionic field induced by a single boundary with
bag boundary condition perpendicular to the string in the Minkowski
spacetime \cite{Saha}. Hence, for a massless field we have the standard
conformal relation between the problems with cosmic strings in AdS spacetime
and in the geometry of the cosmic string on background of Minkowski
spacetime with a boundary at $z=0$, on which the field obeys the MIT bag
boundary condition (see ref. \cite{Saha}).

For a massive field, it is possible to express the series (\ref{Iqy}) in a
form more convenient for the further discussion \cite{Saha}:
\begin{eqnarray}
\mathcal{I}(q,y) &=&\frac{2}{q}\sideset{}{'}{\sum}_{l=0}^{p}(-1)^{l}\cos
\left( \pi l/q\right) e^{y\cos (2\pi l/q)}  \notag \\
&&+\frac{4}{\pi }\cos \left( q\pi /2\right) \int_{0}^{\infty }dx\frac{\sinh
\left( qx\right) \sinh \left( x\right) }{\cosh (2qx)-\cos (q\pi )}e^{-y\cosh
(2x)}\ ,  \label{SumForm}
\end{eqnarray}%
where $p$ is the integer part of $q/2$, $p=[q/2]$, and the prime on the sign
of the sum means that the term $l=0$ should be taken with the coefficient
1/2. In the case $1\leq q<2$, the first term in the right-hand side of (\ref%
{SumForm}) should be omitted. By taking into account that $\mathcal{I}%
(1,y)=e^{y}$, we can see that the contribution of the $l=0$ term in (\ref%
{SumForm}) into $\langle 0|\bar{\psi}\psi |0\rangle $ coincides with the FC
in a pure AdS spacetime when the string is absent, denoted below as $\langle
\bar{\psi}\psi \rangle _{\mathrm{AdS}}$. The remaining part corresponds to
the correction induced by the presence of the cosmic string. So we can
write:
\begin{equation}
\langle \bar{\psi}\psi \rangle _{\mathrm{cs}}=\langle 0|\bar{\psi}\psi
|0\rangle -\langle \bar{\psi}\psi \rangle _{\mathrm{AdS}}\ .  \label{FCcs}
\end{equation}%
For points outside the string the local geometry is the same as in a pure
AdS spacetime and in the absence of a cutoff function the divergent parts of
separate terms in the right-hand side of (\ref{FCcs}) cancel out. Hence, the
string-induced part is finite and does not require any renormalization
procedure.

Substituting (\ref{SumForm}) into (\ref{FC3}) and integrating over the
variable $y$ by using the formula from ref. \cite{Grad}, for the
string-induced part one finds
\begin{eqnarray}
\langle \bar{\psi}\psi \rangle _{\mathrm{cs}} &=&-\frac{\rho ^{-2}}{2\pi
^{2}a^{3}}\bigg[\sum_{l=1}^{p}(-1)^{l}\cos \left( \frac{\pi l}{q}\right)
\frac{W_{ma}(1+2\rho ^{2}\sin ^{2}(\pi l/q))}{\sin ^{2}(\pi l/q)(1+\rho
^{2}\sin ^{2}(\pi l/q))}  \notag \\
&&+\frac{2q}{\pi }\cos \left( \frac{\pi q}{2}\right) \int_{0}^{\infty }\ dx%
\frac{\sinh (qx)\sinh (x)}{\cosh (2qx)-\cos (q\pi )}\frac{W_{ma}(1+2\rho
^{2}\cosh ^{2}x)}{(1+\rho ^{2}\cosh ^{2}x)\cosh ^{2}x}\bigg]\ ,
\label{FCcs2}
\end{eqnarray}%
where%
\begin{equation}
W_{\nu }(u)=Q_{\nu -1}^{2}(u)-Q_{\nu }^{2}(u),  \label{Fnu}
\end{equation}%
and $Q_{\nu }^{\mu }(u)$ is the associated Legendre function of the second
kind. Note that for odd integer values of $q$ the integral term in (\ref%
{FCcs2}) vanishes. As it is seen, the part in the FC\ induced by the cosmic
string depends on the coordinates $r$ and $z$ in the form of the ratio $\rho
$. This is a consequence of the maximal symmetry of AdS spacetime. Noting
that the proper distance from the string is given by $ar/z$, we see that $%
\rho $ is the proper distance of the observation point from the string,
measured in units of the AdS curvature radius $a$.

Let us consider the behavior of $\langle \bar{\psi}\psi \rangle _{\mathrm{cs}%
}$ in some asymptotic regions of the parameters. At large proper distances
from the string, compared with the AdS curvature radius, one has $\rho \gg 1$
and the argument of the associated Legendre functions in (\ref{FCcs2}) is
large. By taking into account that for $u\gg 1$ one has
\begin{equation}
Q_{\nu }^{\mu }(u)\approx \sqrt{\pi }\frac{e^{i\mu \pi }}{2^{\nu +1}}\frac{%
\Gamma (\nu +\mu +1)}{\Gamma (\nu +3/2)}\frac{1}{u^{\nu +1}},  \label{Quas}
\end{equation}%
to leading order we find:%
\begin{equation}
\langle \bar{\psi}\psi \rangle _{\mathrm{cs}}\approx -\frac{(z/r)^{2ma+4}}{%
2^{2ma+1}\pi ^{3/2}a^{3}}\frac{\Gamma (ma+2)}{\Gamma (ma+1/2)}g(q,2ma+4).
\label{FCcsAs}
\end{equation}%
In (\ref{FCcsAs}), we have introduced the notation%
\begin{equation}
g(q,\alpha )=\sum_{l=1}^{p}\frac{(-1)^{l}\cos (\pi l/q)}{\sin ^{\alpha }(\pi
l/q)}+\frac{2q}{\pi }\int_{0}^{\infty }\ dx\frac{\sinh (qx)\sinh (x)}{\cosh
(2qx)-\cos (q\pi )}\frac{\cos (q\pi /2)}{\cosh ^{\alpha }x}.  \label{gq}
\end{equation}%
It can be checked that
\begin{equation}
g(q,4)=-\frac{(q^{2}-1)(7q^{2}+17)}{720}.  \label{gq4}
\end{equation}%
In particular, from (\ref{FCcsAs}) it follows that for a fixed value of $r$
the FC vanishes on the AdS boundary as $z^{2ma+4}$. For a fixed value of $z$
and at large distances from the cosmic string, the string-induced part in
the FC decays as $r^{-2ma-4}$. Note that for a string in background of
Minkowski spacetime and for a massive fermionic field, at large distances
from the string the FC decays exponentially.

At small proper distances from the string, compared with the AdS curvature
scale, we have $\rho \ll 1$ and the argument of the associated Legendre
functions in (\ref{FCcs2}) is close to 1. In this case, we use the following
asymptotic expression for the associated Legendre function:%
\begin{equation}
Q_{\nu }^{\mu }(u)\approx \frac{e^{i\mu \pi }}{2}\Gamma (\mu )\left( \frac{%
u+1}{u-1}\right) ^{\mu /2}\left[ 1+\frac{\nu (1+\nu )}{\mu -1}\frac{1-u}{2}%
\right] .  \label{Qu1as}
\end{equation}%
To the leading order, from (\ref{FCcs2}) one finds%
\begin{equation}
\langle \bar{\psi}\psi \rangle _{\mathrm{cs}}\approx -\frac{m}{2\pi ^{2}}%
\left( \frac{z}{ar}\right) ^{2}g(q,2).  \label{FCcsAs2}
\end{equation}%
From here it follows that, for a fixed distance from the cosmic string, the
string-induced part in the FC diverges on the AdS horizon. Note that for the
geometry of a cosmic string in Minkowski spacetime, near the string one has $%
\langle \bar{\psi}\psi \rangle _{\mathrm{cs}}^{\mathrm{(M)}}\approx
mg(q,2)/(2\pi ^{2}r^{2})$.

Now we consider large values of the curvature radius for AdS spacetime when $%
y$ is fixed, assuming that $ma\gg 1$. By taking into account that in this
case one has $z\approx a+y$, we see that the degree of the associated
Legendre functions in (\ref{FCcs2}) is large whereas the argument is close
to 1. We have the following asymptotic formula \cite{Erde53a}%
\begin{equation}
Q_{\nu }^{\mu }(\cosh (b/\nu ))\approx e^{i\mu \pi }\nu ^{\mu }K_{\mu }(b),
\label{Qu2as}
\end{equation}%
for $\nu \gg 1$. By using this formula and the recurrence relation for the
associated Legendre function from \cite{Abra72} we find%
\begin{equation}
Q_{\nu }^{\mu }(\cosh (b/\nu ))-Q_{\nu -1}^{\mu }(\cosh (b/\nu ))\approx
b\nu ^{\mu -1}e^{i\mu \pi }K_{\mu -1}(b).  \label{Qu3as}
\end{equation}%
With the help of this relation, from (\ref{FCcs2}) for the FC one gets:%
\begin{eqnarray}
\langle \bar{\psi}\psi \rangle _{\mathrm{cs}} &\approx &-\frac{2m^{3}}{\pi
^{2}}\bigg[\sum_{l=1}^{p}(-1)^{l}\left( \frac{\pi l}{q}\right) f_{1}(2mr\sin
(\pi l/q))  \notag \\
&&+\frac{2q}{\pi }\cos \left( \frac{\pi q}{2}\right) \int_{0}^{\infty }\ dx%
\frac{\sinh (qx)\sinh (x)}{\cosh (2qx)-\cos (q\pi )}f_{1}(2mr\cosh x)\bigg],
\label{FClimM}
\end{eqnarray}%
with the notation $f_{\nu }(x)=K_{\nu }(x)/x^{\nu }$. The expression in the
right-hand side coincides with the FC in the geometry of a cosmic string in
Minkowski spacetime \cite{Saha}.

\begin{figure}[tbph]
\begin{center}
\epsfig{figure=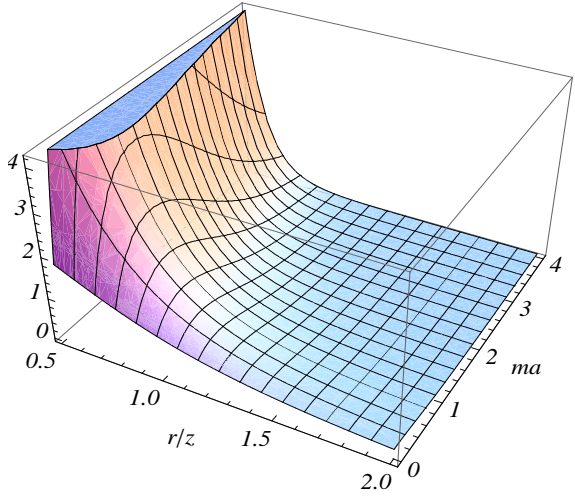,width=7.5cm,height=6.5cm}
\end{center}
\caption{String-induced part in the FC, $a^{3}\langle \bar{\protect\psi}%
\protect\psi \rangle _{\mathrm{cs}}$, as a function of $r/z$ and $ma$ for
the cosmic string with the parameter $q=3$.}
\label{fig1}
\end{figure}

In figure \ref{fig1} we present the dependence of the string-induced part in
the FC on the ratio $r/z$ and on $ma$ for the geometry of cosmic string with
$q=3$. The ratio $r/z$ is the proper distance of the observation point from
the string measured in units of the AdS curvature scale and $ma$ is the
field mass measured in the same units. In the numerical evaluation we have
taken the value $q=3$ in order to simplify the calculations. The qualitative
behaviors for other values of $q$ are similar to the one presented in this
figure.

\section{Energy-momentum tensor}

\label{sec4}

In this section we shall consider another important characteristic of the
fermionic vacuum, the VEV of the energy-momentum tensor. In similar way as
we have used to calculate the FC, this VEV can be evaluated by using the
mode-sum formula below:
\begin{equation}
\langle 0|T_{\mu \nu }|0\rangle =\frac{i}{2}\sum_{\sigma }[\bar{\psi}%
_{\sigma }^{(-)}\gamma _{(\mu }\nabla _{\nu )}\psi _{\sigma }^{(-)}-(\nabla
_{(\mu }\bar{\psi}_{\sigma }^{(-)})\gamma _{\nu )}\psi _{\sigma }^{(-)}]\ ,
\label{modesum}
\end{equation}%
where the brackets enclosing the index mean the symmetrization and $\nabla
_{\mu }\bar{\psi}=\partial _{\mu }\bar{\psi}-\bar{\psi}\Gamma _{\mu }$.
Similar to the case of the FC, the VEV of the energy-momentum tensor is
presented in the decomposed form:
\begin{equation}
\langle 0|T_{\mu \nu }|0\rangle =\langle T_{\mu \nu }\rangle _{\mathrm{AdS}%
}+\langle T_{\mu \nu }\rangle _{\mathrm{cs}}\ ,  \label{EMTdec}
\end{equation}%
where the first and second terms on the right-hand side correspond to the
pure AdS part and to the correction on the VEV of the energy-momentum tensor
due to the presence of the cosmic string. It can be seen that both parts are
diagonal. Because of the maximal symmetry of AdS spacetime, the part $%
\langle T_{\mu \nu }\rangle _{\mathrm{AdS}}$ does not depend on the
spacetime point and is completely determined by its trace: $\langle T_{\mu
\nu }\rangle _{\mathrm{AdS}}=(g_{\mu \nu }/4)\langle T_{\alpha }^{\alpha
}\rangle _{\mathrm{AdS}}$. This part is computed in \cite{Camp1} by using
the zeta function technique and Pauli-Villars regularizations. Again, in
this section we are mainly interested to evaluate the part of the VEV
induced by the cosmic string. This part is finite for points outside the
string and does not require any renormalization procedure.

Let us first evaluate $\langle 0|T_{00}|0\rangle $. Developing the covariant
derivative of the spinor field we observe that, besides the contributions of
the time derivative on the field which provides a term proportional to the
energy, there appear the contribution due to the anti-commutator $\{\gamma
_{0},\ \Gamma _{0}\}$. For this case, the latter vanishes. So, for the VEV
of the energy density one finds
\begin{equation}
\langle 0|T_{00}|0\rangle =\frac{a}{z}\sum_{\sigma }E\ \psi _{\sigma
}^{(-)\dagger}\psi _{\sigma }^{(-)}\ .  \label{T00}
\end{equation}%
Substituting (\ref{psi-}) and (\ref{coef-}) into (\ref{T00}), after the
summation over $s$ we get
\begin{equation}
\langle 0|T_{00}|0\rangle =\frac{qz^{3}}{2\pi a^{2}}\int_{0}^{\infty
}d\lambda \,\lambda \int_{0}^{\infty }dk\,kE\left[ J_{\nu
_{1}}^{2}(kz)+J_{\nu _{2}}^{2}(kz)\right] \sum_{j}\left[ J_{qj+1/2}^{2}(%
\lambda r)+J_{qj-1/2}^{2}(\lambda r)\right] \ .  \label{T00b}
\end{equation}%
Now, to continue this evaluation in a more compact form, we use the identity
\begin{equation}
E=-\frac{2}{\sqrt{\pi }}\int_{0}^{\infty }ds\ \partial _{s^{2}}e^{-\left(
k^{2}+\lambda ^{2}\right) s^{2}}\ .  \label{omRep}
\end{equation}%
Changing the order of integrations, it is possible to integrate over the
variables $\lambda $ and $k$ by using the formula from ref. \cite{Grad}. As
a result, we find the representation
\begin{eqnarray}
\langle 0|T_{00}|0\rangle &=&-\frac{qz^{3}}{4\pi ^{3/2}a^{2}}%
\int_{0}^{\infty }ds\ \partial _{s^{2}}\bigg\{\frac{%
e^{-(r^{2}+z^{2})/(2s^{2})}}{s^{4}}\left[ I_{\nu
_{1}}(z^{2}/(2s^{2}))+I_{\nu _{2}}(z^{2}/(2s^{2}))\right]  \notag \\
&\times &\sum_{j}\left[ I_{qj+1/2}(r^{2}/(2s^{2}))+I_{qj-1/2}(r^{2}/(2s^{2}))%
\right] \bigg\}\ .  \label{T00b1}
\end{eqnarray}%
After the integration by parts over the variable $s$, and using the
summation formula (\ref{SumForm}), the contribution induced by the cosmic
string reads
\begin{eqnarray}
&&\langle T_{0}^{0}\rangle _{\mathrm{cs}}=\frac{\rho ^{-2}}{4\pi ^{2}a^{4}}%
\bigg[\sum_{l=1}^{p}(-1)^{l}\cos \left( \frac{\pi l}{q}\right) \frac{%
F_{ma}^{(2)}(1+2\rho ^{2}\sin ^{2}(\pi l/q))}{\sin ^{2}(\pi l/q)(1+\rho
^{2}\sin ^{2}(\pi l/q))}  \notag \\
&&+\frac{2q}{\pi }\cos \left( \frac{\pi q}{2}\right) \int_{0}^{\infty }\ dx%
\frac{\sinh (qx)\sinh (x)}{\cosh (2qx)-\cos (q\pi )}\frac{%
F_{ma}^{(2)}(1+2\rho ^{2}\cosh ^{2}x)}{(1+\rho ^{2}\cosh ^{2}x)\cosh ^{2}x}%
\bigg]\ ,  \label{T00c}
\end{eqnarray}%
where we have defined a new function
\begin{equation}
F_{\nu }^{(n)}(u)=Q_{\nu -1}^{n}(u)+Q_{\nu }^{n}(u)\ .  \label{Fn}
\end{equation}%
Similar to the FC, the VEV of the energy density depends on $r$ and $z$ in
the form of the combination $\rho $.

For a massless field, by taking into account that $%
F_{0}^{(2)}(1+2y^{2})=1+y^{-2}$ and making use of (\ref{gq4}), from (\ref%
{T00c}) we get:%
\begin{equation}
\langle T_{0}^{0}\rangle _{\mathrm{cs}}=\left( \frac{z}{a}\right)
^{4}\langle T_{0}^{0}\rangle _{\mathrm{cs}}^{\mathrm{(M)}},  \label{T00m0}
\end{equation}%
where%
\begin{equation}
\langle T_{0}^{0}\rangle _{\mathrm{cs}}^{\mathrm{(M)}}=-\frac{%
(q^{2}-1)(7q^{2}+17)}{2880\pi ^{2}r^{4}},  \label{T00M}
\end{equation}%
is the corresponding energy density for a massless field in the geometry of
the cosmic string on Minkowski bulk \cite{ferm}. The massless fermionic
field is conformally invariant and (\ref{T00m0}) presents the standard
relation for the VEVs in conformally related problems.

For large values of $\rho $, $\rho \gg 1$, by making use the asymptotic
expression (\ref{Quas}), to the leading order we find%
\begin{equation}
\langle T_{0}^{0}\rangle _{\mathrm{cs}}\approx -\frac{1}{2a}\langle \bar{\psi%
}\psi \rangle _{\mathrm{cs}}\ ,  \label{T00as}
\end{equation}%
where the asymptotic expression for the FC is given by the expression (\ref%
{FCcsAs}). As it is seen, for a fixed value of $r$, the VEV of the energy
density vanishes on the AdS boundary as $z^{2ma+4}$. At large distances from
the cosmic string, the string-induced part in the energy density behaves as $%
r^{-2ma-4}$. In the opposite limit, $\rho \ll 1$, by using the formula (\ref%
{Qu1as}), we get the following asymptotic expression:%
\begin{equation}
\langle T_{0}^{0}\rangle _{\mathrm{cs}}\approx -\frac{(q^{2}-1)(7q^{2}+17)}{%
2880\pi ^{2}}\left( \frac{z}{ar}\right) ^{4}\ .  \label{T00as1}
\end{equation}%
The string-induced part diverges on the AdS horizon and on the string.

Now let us consider the Minkowskian limit, corresponding to $a\rightarrow
\infty $. By using the relation (\ref{Qu2as}) for the associated Legendre
function, to the leading order one finds
\begin{eqnarray}
&&\langle T_{0}^{0}\rangle _{\mathrm{cs}}\approx \frac{2m^{4}}{\pi ^{2}}%
\bigg[\sum_{l=1}^{p}(-1)^{l}\cos \left( \frac{\pi l}{q}\right) f_{2}(2mr\sin
(\pi l/q))  \notag \\
&&+\frac{2q}{\pi }\cos \left( \frac{\pi q}{2}\right) \int_{0}^{\infty }\ dx%
\frac{\sinh (qx)\sinh (x)}{\cosh (2qx)-\cos (q\pi )}f_{2}(2mr\cosh x)\bigg].
\label{T00as2}
\end{eqnarray}%
The expression in the right-hand side coincides with the corresponding
expression in the geometry of the cosmic string in Minkowski spacetime \cite%
{Saha}.

\begin{figure}[tbph]
\begin{center}
\epsfig{figure=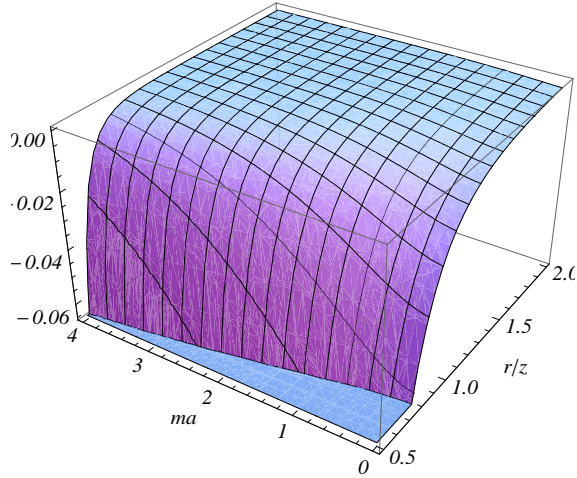,width=7.5cm,height=6.5cm}
\end{center}
\caption{String-induced part in the VEV of the energy density, $a^4 \langle
T_{0}^{0}\rangle _{\mathrm{cs}}$, versus the distance from the string and
the mass of the field in units of AdS curvature scale, for the cosmic string
with $q=3$.}
\label{fig2}
\end{figure}

The dependence of the string-induced part in the VEV of the energy density
on the distance from the string and on the field mass in units of AdS
curvature scale, is displayed in figure \ref{fig2}. As in the numerical
example for the FC, we have considered the cosmic string with $q=3$.

The calculation of $\langle 0|T_{rr}|0\rangle $ also presents, besides the
contribution due to the radial derivative of the field, a contribution
coming from the anticommutator $\{\gamma _{r},\ \Gamma _{r}\}$. In this case
the latter also vanishes, remaining only the contributions due to the radial
derivative,
\begin{equation}
\langle 0|T_{rr}|0\rangle =\frac{i}{2}\sum_{\sigma }[\bar{\psi}_{\sigma
}^{(-)}\gamma _{r}\partial _{r}\psi _{\sigma }^{(-)}-(\partial _{r}\bar{\psi}%
_{\sigma }^{(-)})\gamma _{r}\psi _{\sigma }^{(-)}]\ .  \label{Trr}
\end{equation}%
Substituting (\ref{psi-}) and (\ref{coef-}) into this formula, using the
representation (\ref{gam02}) for the Dirac matrices, after some intermediate
steps we arrive,
\begin{eqnarray}
\langle 0|T_{rr}|0\rangle &=&-\frac{qz^{3}}{8\pi a^{2}}\sum_{\sigma
}\epsilon _{j}\frac{\lambda ^{3}k}{E}\left[ J_{\nu _{1}}^{2}(kz)+J_{\nu
_{2}}(kz)\right]  \notag \\
&&\times \ \left[ J_{\beta _{1}}^{\prime }(\lambda r)J_{\beta _{2}}(\lambda
r)-J_{\beta _{1}}(\lambda r)J_{\beta _{2}}^{\prime }(\lambda r)\right] .
\label{Trr1}
\end{eqnarray}%
Developing the summations over $s$ and $j$ and using the recurrence relation
involving the derivative of the Bessel function, we obtain,
\begin{eqnarray}
\langle 0|T_{rr}|0\rangle &=&\frac{qz^{3}}{2\pi a^{2}}\int_{0}^{\infty
}d\lambda \int_{0}^{\infty }dk\frac{\lambda ^{3}k}{E}\left[ J_{\nu
_{1}}^{2}(kz)+J_{\nu _{2}}^{2}(kz)\right]  \notag \\
&&\times \sum_{j}\left[ J_{qj+1/2}^{2}(\lambda r)+J_{qj-1/2}^{2}(\lambda r)-%
\frac{2qj}{\lambda r}J_{qj-1/2}(\lambda r)J_{qj+1/2}(\lambda r)\right] .
\label{Trr2}
\end{eqnarray}

By using again the identity (\ref{ident1}), the integration over $k$ can be
performed with the help of the formula from ref. \cite{Grad}. The integral
over $\lambda $ requires more details. For the first two terms inside the
brackets, we use the formula below,
\begin{equation}
\int_{0}^{\infty }d\lambda \lambda ^{3}\ e^{-\lambda ^{2}s^{2}}J_{\nu
}^{2}(\lambda r)=-\partial _{s^{2}}\left[ \frac{e^{-r^{2}/(2s^{2})}}{2s^{2}}%
I_{\nu }(r^{2}/(2s^{2}))\right] \ .  \label{Intrel3}
\end{equation}%
As to the integral of the last term, we can adopt a procedure similar to
that used in the obtainment of (\ref{Int1}). The result is,
\begin{eqnarray}
\langle 0|T_{rr}|0\rangle &=&-\frac{qz^{3}}{4\pi ^{3/2}a^{2}}%
\int_{0}^{\infty }ds\frac{e^{-z^{2}/(2s^{2})}}{s^{2}}\left[ I_{\nu
_{1}}(z^{2}/(2s^{2}))+I_{\nu _{2}}(z^{2}/(2s^{2}))\right]  \notag \\
&&\times \sum_{j}\left\{ \partial _{s^{2}}\left[ s^{-2}e^{-r^{2}/(2s^{2})}%
\left( I_{qj+1/2}(r^{2}/(2s^{2}))+I_{qj-1/2}(r^{2}/(2s^{2}))\right) \right]
\right.  \notag \\
&&+\left. \frac{qj}{s^{4}}e^{-r^{2}/(2s^{2})}\left[
I_{qj-1/2}(r^{2}/(2s^{2})-I_{qj+1/2}(r^{2}/(2s^{2})))\right] \right\} \ .
\label{Trra}
\end{eqnarray}%
Finally, introducing a new variable, $y=r^{2}/(2s^{2})$ and using the
relation
\begin{equation}
qj[I_{qj-1/2}(y)-I_{qj+1/2}(y)]=(y\partial
_{y}-y+1/2)[I_{qj-1/2}(y)+I_{qj+1/2}(y)]\ ,  \label{relBes1}
\end{equation}%
all the summations involving the modified Bessel function associated to the
radial coordinate can be developed by using (\ref{SumForm}). Finally
extracting from the general result the contribution due to the pure AdS
spacetime, we obtain the contribution induced by the cosmic string only. It
coincides with the corresponding result for the energy density:
\begin{equation}
\langle T_{r}^{r}\rangle _{\mathrm{cs}}=\langle T_{0}^{0}\rangle _{\mathrm{cs%
}}\ .  \label{Trr00}
\end{equation}

Now let us analyze the azimuthal stress $\langle 0|T_{\phi \phi }|0\rangle $%
. In this case also the only non-vanishing contributions come from the
derivative of the wave-function with respect to the azimuthal variable. This
derivative can be evaluated by using $\partial _{\phi }=iJ_{\phi }-iq\Sigma
^{z}/2$. Taking this procedure into account, there appear an anti-commutator
$\{\gamma _{\phi },\ \Sigma ^{z}\}$, which is zero. So we get,
\begin{equation}
\langle 0|T_{\phi \phi }|0\rangle =q\sum_{\sigma }\ j\ \bar{\psi}_{\sigma
}^{(-)}\gamma _{\phi }\psi _{\sigma }^{(-)}\ .  \label{T22}
\end{equation}%
Substituting (\ref{psi-}) and (\ref{coef-}) into the above equation, and
also using the representation given in (\ref{gam02}) for the Dirac matrices,
we arrive at,
\begin{equation}
\langle 0|T_{\phi \phi }|0\rangle =\frac{q^{2}rz^{3}}{4\pi a^{2}}%
\sum_{\sigma }\epsilon _{j}j\frac{\lambda ^{2}k}{E}J_{\beta _{1}}(\lambda
r)J_{\beta _{2}}(\lambda r)\left[ J_{\nu _{1}}^{2}(kz)+J_{\nu _{2}}^{2}(kz)%
\right] \ .  \label{T221}
\end{equation}%
Because the term inside the summation does not depend on $s$, the sum over
this quantum number provides only a factor $2$. As to the summation over $j$
we have
\begin{equation}
\sum_{j=\pm 1/2,\cdots }j\epsilon _{j}J_{\beta _{1}}(\lambda r)J_{\beta
_{2}}(\lambda r)=2\sum_{j}jJ_{qj-1/2}(\lambda r)J_{qj+1/2}(\lambda r)\ .
\label{Sumj22}
\end{equation}%
By using the relation (\ref{ident1}), the integral over the variable $k$ can
be performed with the help of the formula from ref. \cite{Grad}. As to the
integral over $\lambda $, we can use similar procedure as presented in (\ref%
{Int1}). Defining the new variable $y=r^{2}/(2s^{2})$, we obtain
\begin{eqnarray}
\langle 0|T_{\phi \phi }|0\rangle &=&\frac{2q\rho ^{-3}}{(2\pi )^{3/2}a^{2}}%
\int_{0}^{\infty }dy\ y^{3/2}e^{-(1+\rho ^{-2})y}\left[ I_{\nu _{1}}(y/\rho
^{2})+I_{\nu _{2}}(y/\rho ^{2})\right]  \notag \\
&\times &\sum_{j}qj\left[ I_{qj-1/2}(y)-I_{qj+1/2}(y)\right] \ .
\label{Tppb}
\end{eqnarray}%
Using (\ref{relBes1}), it is possible to express the azimuthal stress in
terms of the function (\ref{Iqy}). By using formula (\ref{SumForm}), the
integral over the variable $y$ can be performed explicitly. As a result, the
contribution to the azimuthal stress induced by the cosmic string reads,
\begin{eqnarray}
\langle T_{\phi }^{\phi }\rangle _{\mathrm{cs}} &=&\frac{a^{-4}}{4\pi ^{2}}%
\bigg[\sum_{l=1}^{p}(-1)^{l}\cos \left( \frac{\pi l}{q}\right) G_{ma}(\rho
\sin (\pi l/q))  \notag \\
&&+\frac{2q}{\pi }\cos \left( \frac{q\pi }{2}\right) \int_{0}^{\infty }\ dx%
\frac{\sinh (qx)\sinh (x)G_{ma}(\rho \cosh x)}{\cosh (2qx)-\cos (q\pi )}%
\bigg]\ ,  \label{Tppc}
\end{eqnarray}%
where%
\begin{equation}
G_{\nu }(x)=\frac{F_{\nu }^{(2)}(1+2x^{2})}{x^{2}(1+x^{2})}+2\frac{F_{\nu
}^{(3)}(1+2x^{2})}{x(1+x^{2})^{3/2}}.  \label{Gnux}
\end{equation}%
As for the other components, the azimuthal stress depends on $r$ and $z$ in
the form of the ratio $\rho $. The latter is the proper distance from the
string measured in units of the AdS curvature radius. For a massless
fermionic field one has $G_{0}(x)=-3/x^{4}$, and from (\ref{Tppc}) one gets $%
\langle T_{\phi }^{\phi }\rangle _{\mathrm{cs}}=-3\langle T_{0}^{0}\rangle _{%
\mathrm{cs}}$, where the $\langle T_{0}^{0}\rangle _{\mathrm{cs}}$ is given
by (\ref{T00m0}).

Let us consider the behavior of the azimuthal stress in the asymptotic
regions of the ratio $\rho $. For $\rho \gg 1$ we use the formula (\ref{Quas}%
) for the associated Legendre function. To the leading order this gives:%
\begin{equation}
\langle T_{\phi }^{\phi }\rangle _{\mathrm{cs}}\approx \frac{2ma+3}{2a}%
\langle \bar{\psi}\psi \rangle _{\mathrm{cs}}.  \label{T33as1}
\end{equation}%
Comparing with (\ref{T00as}), we see that in the region under consideration
one has the relation $\langle T_{\phi }^{\phi }\rangle _{\mathrm{cs}}\approx
-(2ma+3)\langle T_{0}^{0}\rangle _{\mathrm{cs}}$. For $\rho \ll 1$, similar
to the case of the energy density, we get the following relation%
\begin{equation}
\langle T_{\phi }^{\phi }\rangle _{\mathrm{cs}}\approx -3\langle
T_{0}^{0}\rangle _{\mathrm{cs}},  \label{T33as2}
\end{equation}%
where the asymptotic expression for the energy density is given by (\ref%
{T00as1}). And finally, in the Minkowskian limit, corresponding to $%
a\rightarrow \infty $, we recover the result from \cite{Saha}:
\begin{eqnarray}
\langle T_{\phi }^{\phi }\rangle _{\mathrm{cs}} &\approx &-\frac{2m^{4}}{\pi
^{2}}\left[ \sum_{l=1}^{p}(-1)^{l}\cos (\pi l/q)f^{(\phi )}(2mr\sin (\pi
l/q))\right.  \notag \\
&&+\left. \frac{2q}{\pi }\cos \left( \frac{q\pi }{2}\right) \int_{0}^{\infty
}\ dx\frac{\sinh (qx)\sinh (x)f^{(\phi )}(2mr\cosh x)}{\cosh (2qx)-\cos
(q\pi )}\right] \ .  \label{T33as3}
\end{eqnarray}%
with%
\begin{equation}
f^{(\phi )}(x)=f_{1}(x)+3f_{2}(x).  \label{fphi}
\end{equation}

\begin{figure}[tbph]
\begin{center}
\epsfig{figure=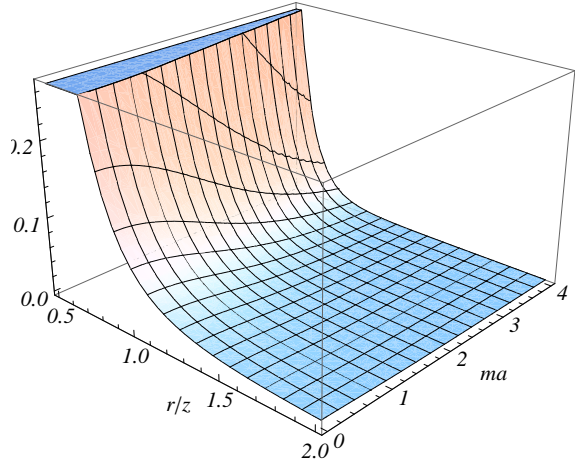,width=7.5cm,height=6.5cm}
\end{center}
\caption{String-induced part in the VEV of the azimuthal stress, $a^4
\langle T_{\protect\phi}^{\protect\phi}\rangle _{\mathrm{cs}}$, as a
functions of the distance from the string and of the field mass in units of
AdS curvature scale for the cosmic string with $q=3$.}
\label{fig3}
\end{figure}

Figure \ref{fig3} presents the string-induced part in the VEV of the
azimuthal stress as a function of $r/z$ and $ma$ (proper distance from the
string and the mass measured in the units of the AdS curvature scale $a$).
We have considered the cosmic string with $q=3$.

The calculation of the axial stress, $\langle 0|T_{zz}|0\rangle $, becomes
simpler because $\Gamma _{z}=0$. Substituting (\ref{psi-}) and (\ref{coef-}%
), and the corresponding gamma matrix into (\ref{modesum}), after some
intermediate steps we arrive at:
\begin{eqnarray}
\langle 0|T_{zz}|0\rangle &=&-\frac{qz^{3}}{8\pi a^{2}}\sum_{\sigma }\frac{%
\lambda ^{2}k^{2}}{E}\left[ b_{s}^{(+)}J_{\beta _{1}}^{2}(\lambda
r)-b_{s}^{(-)}J_{\beta _{2}}^{2}(\lambda r)\right]  \notag \\
&&\times \left[ J_{\nu _{1}}(kz)J_{\nu _{2}}^{\prime }(kz)-J_{\nu
_{2}}(kz)J_{\nu _{1}}^{\prime }(kz)\right] \ .  \label{Tzza}
\end{eqnarray}%
By using the recurrence relations involving the derivative of the Bessel
function, taking the summations over $s$, and using (\ref{ident1}) we get:
\begin{eqnarray}
\langle 0|T_{zz}|0\rangle &=&\frac{qz^{3}}{\pi ^{3/2}a^{2}}\int_{0}^{\infty
}ds\int_{0}^{\infty }d\lambda \lambda e^{-\lambda ^{2}s^{2}}\sum_{j}\left[
J_{qj-1/2}^{2}(\lambda r)+J_{qj+1/2}^{2}(\lambda r)\right]  \notag \\
&&\int_{0}^{\infty }dkk^{3}e^{-k^{2}s^{2}}\left[ J_{\nu _{1}}^{2}(kz)+J_{\nu
_{2}}^{2}(kz)-\frac{2ma}{kz}J_{\nu _{1}}(kz)J_{\nu _{2}}(kz)\right] \ .
\label{Tzza2}
\end{eqnarray}%
The procedure to evaluate the integrals is similar to the one we have
adopted in the calculation of the radial stress. Introducing in the final
result the new variable $y=r^{2}/(2s^{2})$, we obtain
\begin{eqnarray}
\langle 0|T_{z}^{z}|0\rangle &=&\frac{q\rho ^{-5}}{(2\pi )^{3/2}a^{4}}%
\int_{0}^{\infty }dy\ y^{3/2}e^{-(1+\rho ^{-2})y}\mathcal{I}(q,y)  \notag \\
&&\times \left[ I_{\nu _{1}}(y/\rho ^{2})+I_{\nu _{2}}(y/\rho ^{2})\right] \
.  \label{Tzzb}
\end{eqnarray}%
By using (\ref{SumForm}) and extracting the part due to the pure AdS
spacetime, we obtain the contribution to the axial stress induced by the
cosmic string. The latter coincides with the corresponding expression for
the energy density:
\begin{equation}
\langle T_{z}^{z}\rangle _{\mathrm{cs}}=\langle T_{0}^{0}\rangle _{\mathrm{cs%
}}\ .  \label{Tzz00}
\end{equation}

By using the recurrence relations involving the derivative of the associated
Legendre function, it can be explicitly checked that the part in the VEV of
the energy-momentum tensor induced by the cosmic string obeys the covariant
conservation equation, $\langle T_{\mu }^{\nu }\rangle _{\mathrm{cs};\nu }=0$%
, which for the problem under consideration is reduced to two differential
equations,
\begin{eqnarray}
&&\partial _{r}(r\langle T_{r}^{r}\rangle _{\mathrm{cs}})-\langle T_{\phi
}^{\phi }\rangle _{\mathrm{cs}}=0\ ,  \notag \\
&&z\partial _{z}\langle T_{z}^{z}\rangle _{\mathrm{cs}}+\langle T_{\mu
}^{\mu }\rangle _{\mathrm{cs}}-4\langle T_{z}^{z}\rangle _{\mathrm{cs}}=0\ .
\label{Cons}
\end{eqnarray}%
Also, by using the recurrence relations involving the associated Legendre
functions of different orders, it can be verified that the trace relation
below is satisfied:
\begin{equation}
\langle T_{\mu }^{\mu }\rangle _{\mathrm{cs}}=m\langle \bar{\psi}\psi
\rangle _{\mathrm{cs}}\ .  \label{Trace}
\end{equation}%
Note that the trace anomaly is contained in the pure AdS part of the VEV and
the string-induced part is traceless for a massless fermionic field.

It is of interest to compare the results given above for the VEV of the
energy-momentum tensor with the corresponding results for a scalar field
discussed in \cite{String-AdS}. In the case of the scalar field, the vacuum
energy--momentum tensor, in general, is non-diagonal with the off-diagonal
component $\langle T_{z}^{r}\rangle _{\mathrm{cs}}$. The latter vanishes for
a conformally coupled massless field only. In this special case, the VEV of
the energy-momentum tensor is conformally related to the corresponding
quantity for the geometry of a cosmic string in flat spacetime with an
additional flat boundary with Dirichlet boundary condition on it. Another
difference in the VEVs for scalar and fermionic fields is that, in the
scalar case the radial and axial stresses, in general, do not coincide with
the energy density (no summation over $l$): $\langle T_{l}^{l}\rangle _{%
\mathrm{cs}}\neq \langle T_{0}^{0}\rangle _{\mathrm{cs}}$, $l=r,z$. At large
distances from the string, for a scalar field the diagonal components of the
VEV decay as $\langle T_{\mu }^{\mu }\rangle _{\mathrm{cs}}\propto
(z/r)^{2\nu +3}$, where $\nu =\sqrt{9/4-12\xi +m^{2}a^{2}}$ and $\xi $ is
the curvature coupling parameter. In particular, for a massive conformally
coupled scalar field ($\xi =1/6$) the suppression of the VEVs is weaker than
in the fermionic case with the same mass.

\section{Conclusion}

\label{conc}

In this paper we have evaluated the FC and the VEV of the energy-momentum
tensor associated with fermionic field on background of a four-dimensional
AdS spacetime in the presence of a cosmic string. These VEVs are generated
by the two sources of the vacuum polarization: by the gravitational field
due to the negative cosmological constant and by non-trivial topology
induced by the cosmic string. Because the analysis of quantum fermionic
fields in a pure AdS space have been developed in the literature, here we
are mainly interested in the calculation of the vacuum expectations values
induced by the cosmic string. Moreover, because the presence of the string
does not modify the curvature of the AdS background, all the divergences
presented in the calculations of the VEVs, appear only in the contributions
due the purely AdS space. So the contributions induced by the string do not
require renormalization. All of them are automatically finite for points
outside the string.

The evaluation of the above mentioned VEVs have been made by using the
summation over the fermionic modes. So a crucial point in this paper was to
obtain the complete set of fermionic wave-functions given by (\ref{psi+})
and (\ref{psi-}). In order to specify uniquely the mode-functions in AdS
bulk, an additional boundary condition is required on the AdS boundary.
Here, we consider a special case of boundary conditions, when the MIT bag
boundary condition is imposed at a finite distance from the boundary, which
is then taken to zero. By applying to the sum over the angular momentum the
summation formula (\ref{SumForm}), the VEVs are decomposed as the sums of
the pure AdS background and string-induced parts. In this way, for points
away from the string, the renormalization is reduced to the one for the pure
AdS bulk in the absence of the string. String-induced parts in both the FC
and the VEV of the energy-momentum tensor depend on the coordinates $r$ and $%
z$ in the form of the combination $r/z$. The latter is the proper distance
from the string measured in the units of AdS curvature radius. As partial
check of the formulas for AdS bulk, we have shown that for large values of
the curvature radius, to leading order, the VEVs are obtained for the
geometry of cosmic string in Minkowski spacetime.

The string-induced part in the FC is given by the expression (\ref{FCcs2}).
At large proper distances from the string, the leading term in the
corresponding asymptotic expansion is given by (\ref{FCcsAs}). For a fixed
value of the radial coordinate $r$, the string-induced part in the FC
vanishes on the AdS boundary. For a fixed value of $z$ and at large
distances from the cosmic string, the string-induced part decays as $%
r^{-2ma-4}$. For a cosmic string in background of Minkowski spacetime and
for a massive fermionic field, at large distances from the string the FC
decays exponentially. At small proper distances from the string, the leading
term in FC behaves as $(z/r)^{2}$. In particular, the FC diverges on the AdS
horizon.

The VEV of the energy-momentum tensor is decomposed as (\ref{EMTdec}).
Because the maximal symmetry of the AdS bulk, the pure AdS part does not
depend on the spacetime point and is proportional to the metric tensor. The
string-induced part in the VEV of the energy-momentum tensor is diagonal and
the corresponding axial and radial stresses are equal to the energy density.
Note that, in the case of the scalar field, the vacuum energy--momentum
tensor, in general, is non-diagonal with the off-diagonal component $\langle
T_{z}^{r}\rangle _{\mathrm{cs}}$. Another difference in the VEVs for scalar
and fermionic fields is that, in the scalar case the radial and axial
stresses, in general, do not coincide with the energy density. For the
fermionic field, the string-induced parts in the energy density and the
azimuthal stress are given by the expressions (\ref{T00c}) and (\ref{Tppc}).
At large proper distances from the string, these parts are related to the FC
by formulas (\ref{T00as}) and (\ref{T33as1}) and the total VEV is dominated
by the pure AdS part. At small proper distances from the string, the
string-induced part behaves as $(z/r)^{4}$ and between the energy density
and the azimuthal stress one has the relation (\ref{T33as2}). We have
explicitly checked that the components of the string-induced part of the
energy-momentum tensor obey the covariant conservation conditions, (\ref%
{Cons}), and the trace relation, Eq. (\ref{Trace}). The trace anomaly is
contained in the pure AdS VEV and the string-induced part is traceless for a
massless fermionic field.

In the geometry under consideration the boundary of AdS spacetime can be
identified with a 3-dimensional conical spacetime. As it has been shown in
\cite{Dese84}, this spacetime arises as a solution of 3-dimensional Einstein
equations in the presence of a point mass $m_{0}$. This mass is connected to
the planar angle deficit by the relation $m_{0}=(1-1/q)/(4G_{3})$, where $%
G_{D}$ is the gravitational constant in $D$-dimensional spacetime.
Note that the linear mass density of the string in 4-dimensional
spacetime is given by $\mu _{0}=(1-1/q)/(4G_{4})$. Similar to
AdS/CFT correspondence, a duality between the scalar field
theories living in AdS bulk with an infinite static string and on
its boundary has been discussed in \cite{Cristine}. Analogous
duality between the fermionic field theories can be considered by
using the
standard prescription for fermion fields in AdS/CFT correspondence \cite%
{Henn98,Muec98} (for a recent discussion see also \cite{Iqba09}).
In this prescription the non-normalizable modes in AdS bulk are
treated as source terms coupled to a fermionic operator
$\mathcal{O}$ in the boundary theory which is dual to the
fermionic field $\psi $. The quantization procedure we have used
corresponds to, so called, standard quantization of fermionic
fields in AdS/CFT correspondence. For $ma<1/2$ there is an
alternative quantization procedure in which the boundary condition
fixes the lower component of the bispinor. In this case one has
two different dual theories on the boundary which are related by a
Legendre transform. Having the expectation values for the bilinear
products of the fermionic operators in the bulk one can
investigate the corresponding expectation values in the dual
theory by using the standard prescription for computing the
two-point functions on the boundary. However, such a discussion is
beyond the scope of the present paper.

\section*{Acknowledgment}

E.R.B.M. thanks Conselho Nacional de Desenvolvimento Cient\'{\i}fico e Tecnol%
\'{o}gico (CNPq) for partial financial support. A.A.S. was supported by CNPq.

\appendix

\section{On the boundary condition at AdS boundary}

Consider the fermionic field in the region $z>\delta $ assuming that on the
boundary the field obeys the MIT bag boundary condition
\begin{equation}
(1+i\gamma ^{\mu }n_{\mu })\psi =0,\;z=\delta ,  \label{Bagbc}
\end{equation}%
with $n_{\mu }=-\delta _{\mu }^{z}a/z$ being the normal to the boundary.
This boundary condition guarantees the zero flux of fermions through the
boundary. The positive-energy mode functions for this problem are obtained
from (\ref{psi+}) with the replacements $C_{\sigma }^{(+)}\rightarrow
C_{b\sigma }^{(+)}$ and $J_{\nu _{l}}(kz)\rightarrow J_{\nu
_{l}}(kz)+CY_{\nu _{l}}(kz)$, $l=1,2$. From the boundary condition (\ref%
{Bagbc}) it follows that $C=-J_{\nu _{2}}(k\delta )/Y_{\nu _{2}}(k\delta )$.
The new normalization coefficient $C_{b\sigma }^{(+)}$ is determined from
the normalization condition (\ref{normcond}), where now the integration goes
over the region $z\geqslant \delta $. By the calculations similar to those
we have described in section \ref{sec2}, the following expression is
obtained
\begin{equation}
|C_{b\sigma }^{(+)}|^{2}=\frac{sq\lambda ^{2}k}{8\pi a^{3}Eb_{s}^{(+)}}\frac{%
Y_{\nu _{2}}^{2}(k\delta )}{J_{\nu _{2}}^{2}(k\delta )+Y_{\nu
_{2}}^{2}(k\delta )}.  \label{Cbsig}
\end{equation}

In this way, for the positive-energy fermionic mode functions, obeying the
boundary condition (\ref{Bagbc}), one gets%
\begin{equation}
\psi _{b\sigma }^{(+)}(x)=C_{b\sigma }^{(+)}e^{-iEt}z^{2}\left(
\begin{array}{c}
J_{\beta _{1}}(\lambda r)V_{\nu _{1}}(kz) \\
-\epsilon _{j}b_{s}^{(+)}J_{\beta _{2}}(\lambda r)V_{\nu _{2}}(kz)e^{iq\phi }
\\
isJ_{\beta _{1}}(\lambda r)V_{\nu _{2}}(kz) \\
i\epsilon _{j}sb_{s}^{(+)}J_{\beta _{2}}(\lambda r)V_{\nu _{1}}(kz)e^{iq\phi
}%
\end{array}%
\right) e^{iq(j-1/2)\phi }\ ,  \label{psi+b}
\end{equation}%
with the notation
\begin{equation}
V_{\nu }(kz)=J_{\nu }(kz)-\frac{J_{\nu _{2}}(k\delta )}{Y_{\nu _{2}}(k\delta
)}Y_{\nu }(kz).  \label{Vnu}
\end{equation}%
The expression for the negative-energy modes $\psi _{b\sigma }^{(-)}(x)$ is
derived in a similar way. This expression is obtained from (\ref{psi-}) by
the replacements $C_{\sigma }^{(-)}\rightarrow C_{b\sigma }^{(-)}$ and $%
J_{\nu _{l}}(kz)\rightarrow V_{\nu _{l}}(kz)$, where $|C_{b\sigma
}^{(-)}|^{2}$ is given by (\ref{Cbsig}) with the change $b_{s}^{(+)}%
\rightarrow b_{s}^{(-)}$. In the limit $\delta \rightarrow 0$, for a fixed $%
z>\delta $, one has $|C_{b\sigma }^{(+)}|^{2}\rightarrow |C_{\sigma
}^{(+)}|^{2}$, $V_{\nu _{l}}(kz)\rightarrow J_{\nu _{l}}(kz)$ and the mode
functions $\psi _{b\sigma }^{(\pm )}(x)$ are reduced to (\ref{psi+}) and (%
\ref{psi-}), as it has been stated in the text.

\end{document}